\documentclass[a4paper,12pt, epsfig]{article}\def\letter{0}\def\pr{0}

\usepackage{epsfig}
\usepackage{epstopdf}
\usepackage{graphicx}
\usepackage{ifthen}

\usepackage{amsmath}
\usepackage[psamsfonts]{amssymb}
\usepackage{euscript}

\usepackage{latexsym}
\usepackage[arrow,matrix,curve]{xy}

\newskip\humongous \humongous=0pt plus 1000pt minus 1000pt

\newif\ifdtup

\def\,{\hspace{-.1cm}}
\def\hsp{,\hspace{.7cm}}

\def\tg{{\tilde{g}}}

\def\op{\omega_p}

\def\fc#1#2 {\frac{n}{q}#1\frac{n}{q}#2}

\newcommand{\vac}{\ensuremath{|0\rangle}}

\renewcommand{\tanh}{\textrm{tanh}}
\newcommand{\sech}{\textrm{sech}}

\renewcommand{\theequation}{\arabic{section}.\arabic{equation}}
\renewcommand{\(}{\begin{equation}}
\renewcommand{\)}{end{equation} \vspace{-.05in}\linebreak}

\newcounter{saveeqn}
\newcounter{savealpheqn}

\newcommand{\alpheqn}{\setcounter{saveeqn}{\value{equation}}%
  \stepcounter{saveeqn}\setcounter{equation}{0}%
  \renewcommand{\theequation}{\mbox{\arabic{section}.\arabic{saveeqn}
\alph{equation}}}
  \renewcommand{\)}{\end{equation}}}
\def\part#1{\frac{\partial}{\partial{#1}}}%
\def\group#1{\refstepcounter{equation}\setcounter{saveeqn}
 {\value{equation}}%
  \label{#1}\setcounter{equation}{0}%
\renewcommand{\theequation}{\mbox{\arabic{section}.\arabic{saveeqn}
\alph{equation}}}
  \renewcommand{\)}{\end{equation}}}
\newcommand{\reseteqn}{\setcounter{equation}{\value{saveeqn}}%
  \renewcommand{\theequation}{\arabic{section}.\arabic{equation}}%
  \renewcommand{\)}{\end{equation}}}

\newcommand{\aalpheqn}{\setcounter{saveeqn}{\value{equation}}%
  \stepcounter{saveeqn}\setcounter{equation}{0}%
  \renewcommand{\theequation}{\mbox{
        \Alph{subsection}.\arabic{saveeqn}\alph{equation}}}
   \renewcommand{\)}{\end{equation}}}
\newcommand{\areseteqn}{\setcounter{equation}{\value{saveeqn}}%
  \renewcommand{\theequation}{\Alph{subsection}.\arabic{equation}}%
  \renewcommand{\)}{\end{equation}}}

\renewcommand{\thefootnote}{\alph{footnote}}
\renewcommand{\(}{\begin{equation}}
\renewcommand{\)}{\end{equation}}
\newcommand{\ba}{\begin{eqnarray}}
\newcommand{\ea}{\end{eqnarray}}
\newcommand{\cbp}{\mathop{\vtop{\ialign{##\crcr
   $\hfil\displaystyle{}\hfil$\crcr\noalign{\kern-13pt\nointerlineskip}
   \BIG{)}\hskip0pt\crcr\noalign{\kern3pt}}}}}
\newcommand{\pa}{\mathop{\vtop{\ialign{##\crcr

$\hfil\displaystyle{\oplus}\hfil$\crcr\noalign{\kern+1pt\nointerlineskip
}
   \hspace{.08in}$^{\alpha=0}$\hskip6pt\crcr\noalign{\kern3pt}}}}}
\renewcommand{\hsp}{,\hspace{.3in}}
\newcommand{\p}{^\prime}

\def\rv{\rho^{\rm{(vac)}}}

\catcode`\@=11
\def\vereq#1#2{\lower3pt\vbox{\baselineskip1.5pt \lineskip1.5pt
\ialign{$\m@th#1\hfill##\hfil$\crcr#2\crcr\sim\crcr}}}
\catcode`\@=12

\renewcommand{\(}{\begin{equation}}
\renewcommand{\)}{\end{equation}}

\def\pin#1{\int \frac{d#1}{2\pi}}
\def\pink#1{\int \frac{d^{#1}k}{(2\pi)^{#1}}}

\def\dpin#1{\int^+ \frac{d#1}{2\pi}}
\def\dpink#1{\int^+ \frac{d^{#1}k}{(2\pi)^{#1}}}

\def\df{\mathcal{D}_f}

\def\I{\mathcal{I}}
\def\vl#1#2#3{\vector(#1,#2){#3}\line(#1,#2){#3}}

\newcommand{\beas}{\begin{eqnarray*}}
\newcommand{\eeas}{\end{eqnarray*}}

\newcommand{\bquo}{\begin{quote}}
\newcommand{\enqu}{\end{quote}}

\def\ch{{\mathcal{H}}}
\def\chct{{\mathcal{H}_{\rm{c.t.}}}}

\def\ok#1{\omega_{k_{#1}}}

\def\V#1{V^{(#1)}[\sqrt{\lambda}f(x)]}

\def\kkk{\sum_{i}^3k_i}

\newcommand{\beq}{\begin{equation}}
\newcommand{\eeq}{\end{equation}}
\newcommand{\bea}{\begin{eqnarray}}
\newcommand{\eea}{\end{eqnarray}}

\newskip\humongous \humongous=0pt plus 1000pt minus 1000pt

\newif\ifdtup

\catcode`\@=11

\ifthenelse{\equal{\letter}{0}}{ 

\@addtoreset{equation}{section}
\def\theequation{\arabic{section}.\arabic{equation}}

\def\@normalsize{\@setsize\normalsize{15pt}\xiipt\@xiipt
\abovedisplayskip 14pt plus3pt minus3pt%
\belowdisplayskip \abovedisplayskip
\abovedisplayshortskip \z@ plus3pt%
\belowdisplayshortskip 7pt plus3.5pt minus0pt}

\def\small{\@setsize\small{13.6pt}\xipt\@xipt
\abovedisplayskip 13pt plus3pt minus3pt%
\belowdisplayskip \abovedisplayskip
\abovedisplayshortskip \z@ plus3pt%
\belowdisplayshortskip 7pt plus3.5pt minus0pt
\def\@listi{\parsep 4.5pt plus 2pt minus 1pt
      \itemsep \parsep
      \topsep 9pt plus 3pt minus 3pt}}

\relax

\def\section{\@startsection{section}{1}{\z@}{3.5ex plus 1ex minus  .2ex}{2.3ex plus .2ex}{\large\bf}}

\def\thesection{\arabic{section}}
\def\thesubsection{\arabic{section}.\arabic{subsection}}

\def\appendix{\setcounter{section}{0}
 \def\thesection{Appendix \Alph{section}}
 \def\thesubsection{\Alph{section}.\arabic{subsection}}
 \def\theequation{\Alph{section}.\arabic{equation}}}
\renewcommand{\theequation}{\arabic{section}.\arabic{equation}}

}{
\renewcommand{\theequation}{\arabic{equation}}

} 

\begin{document}
\def\thefootnote{\fnsymbol{footnote}}
\def\thetitle{The Two-Loop $\phi^4$ Kink Mass}
\def\autone{Jarah Evslin}
\def\affa{Institute of Modern Physics, NanChangLu 509, Lanzhou 730000, China}
\def\affb{University of the Chinese Academy of Sciences, YuQuanLu 19A, Beijing 100049, China}

\title{Titolo}

\ifthenelse{\equal{\pr}{1}}{
\title{\thetitle}
\author{\autone}
\affiliation {\affa}
\affiliation {\affb}

}{

\begin{center}
{\large {\bf \thetitle}}

\bigskip

\bigskip


{\large \noindent  \autone{${}^{1,2}$}}


\vskip.7cm

1) \affa\\
2) \affb\\

\end{center}

}

\begin{abstract}
\noindent
At one loop, quantum kinks are described by a free theory.  The nonlinearity and so the interesting phenomenology arrives at two loops, where, for example, internal excitations couple to continuum excitations.  We calculate the two-loop mass of a scalar kink.    Unlike previous calculations, we include a counterterm which cancels the vacuum energy density at this order, so that our result for the kink mass is finite even when the vacuum energy density is nonzero.   This allows us to apply our result to the $\phi^4$ kink, for which we obtain a two-loop mass contribution of $0.0126\lambda/m$ in terms of the coupling $\lambda$ and the meson mass $m$ evaluated at the minimum of the potential.

\noindent

\end{abstract}

%
\setcounter{footnote}{0}
\renewcommand{\thefootnote}{\arabic{footnote}}

\ifthenelse{\equal{\pr}{1}}{
\maketitle
}{}

\section{Background}

The $\phi^4$ double well model, and its kink solution, have infiltrated many fields of physics.  For example, this kink appeared in long-chain polyenes in Ref.~\cite{ssh}, a discovery which led to the 2000 Nobel Prize in Chemistry for one of its authors.  This built upon the study of $\phi^4$ kinks as defects in crystals \cite{crystal}.  They have also been observed in graphene nanoribbons \cite{graphene} together with a rich sampling of long-predicted kink phenomenology.

Modern treatments of the quantum $\phi^4$ kink can trace their origins to Ref.~\cite{dhn2}, which found the first quantum correction to the kink mass, arising at one loop.  At one loop the mass is determined by a free truncation of the theory, and so it only depends on the density of states \cite{gj}.  However, it has been appreciated since Ref.~\cite{crystal} that the kinks become interesting, and useful, at two loops, where the normal modes begin to interact nonlinearly.   Ref.~\cite{vega} began a calculation of the two-loop mass, showing that the result is finite, but no answer was presented.  The various mass contributions at two loops also appeared, in a more general setting, in Ref.~\cite{sl}.

In the 45 years since Ref.~\cite{dhn2}, the quantum kink mass has been calculated using lattice techniques \cite{lat98,latt09}, truncated Hamiltonian techniques in the instant frame \cite{mussardo,slava,bajnok} and the light front frame \cite{vary03} as well as Borel resummation \cite{serone}.  In each case, the difference between the calculated mass and the classic result of Ref.~\cite{dhn2} was essentially of the same order as the uncertainty in the calculation, except for Ref.~\cite{latt09} whose results are in strong tension with the majority of the other studies.   In this note we will present a calculation of the two-loop correction to the kink mass with a fractional uncertainty of less than $10^{-3}$.

The two-loop energy $Q_2$ of the ground state of a kink in a general theory with Hamiltonian
\bea
H&=&\int dx \ch(x) \label{hd}\\
\ch(x)&=&\frac{1}{2}:\pi(x)\pi(x):_a+\frac{1}{2}:\partial_x\phi(x)\partial_x\phi(x):_a\nonumber\\
&&+\frac{1}{\lambda}:V[\sqrt{\lambda}\phi(x)]:_a\nonumber
\eea
and classical kink solution
\beq
\phi(x,t)=f(x) \label{fd}
\eeq
was calculated in Ref.~\cite{me2looplett}. Here  $::_a$ is normal ordering of the operators that create and annihilate plane waves.  These operators and indeed all states and operators in this note are defined in the Schrodinger picture.    

Let $V^{(n)}$ be the $n$th derivative of $\lambda^{n/2-1}V[\sqrt{\lambda}\phi]$ with respect to $\sqrt{\lambda}\phi$.  If $V^{(3)}[\sqrt{\lambda}f(\pm\infty)]$ is nonzero then $Q_2$ will be infinite.  This infinity is to be expected, as in this case the vacuum has a finite energy density $\rv$ and so an infinite energy.  The kink mass is intuitively the difference between these two infinities.  The main result of this note is a calculation of this kink mass.

\section{Review of the Kink Energy}

First let us summarize the derivation of the kink energy $Q_2$, then we will modify it to obtain the kink mass.

Let $|K\rangle$ be the kink ground state. It is a Hamiltonian eigenstate
\beq
H|K\rangle=Q|K\rangle.
\eeq
Define the unitary displacement operator
\beq
\df={\rm{exp}}\left(-i\int dx f(x)\pi(x)\right) \label{df}
\eeq
which creates a kink from the vacuum, although not in a Hamiltonian eigenstate.  Then the state
\beq
\vac=\df^\dag |K\rangle
\eeq
is an eigenstate of the kink Hamiltonian $H\p$
\beq
H\p\vac=Q\vac\hsp H\p=\df^\dag H\df. \label{hp}
\eeq
We will expand $\vac$ and $Q$ in powers of $\hbar^{1/2}$ and $\hbar$ respectively
\beq
\vac=\sum_{i=0}\vac_i\hsp
Q=\sum_{j=0}Q_j. \label{semi}
\eeq
$Q_j$ and $\vac_{2j-2}$ are determined at $j$-loops.

 Define the normal modes $g_k(x)$ to be the orthonormal classical solutions 
\bea
\phi(x,t)&=&e^{i\omega_k t}g_k(x)\hsp \omega_k=\sqrt{m^2+k^2}\nonumber\\
 g_k(-x)&=&g_k^*(x)=g_{-k}(x)
\eea
of the linearized equations of motion derived from $H\p$.  Here $m^2=V^{(2)}[\sqrt{\lambda}f(\pm\infty)]$.  If these two limits are not equal, then the kink will experience a force \cite{force1,force2} and so will not be a Hamiltonian eigenstate, and so its mass will not be defined.  There will always be continuum normal modes, corresponding to all real values of $k$.  There will also be discrete bound modes, corresponding to imaginary values of $k$, including a zero mode $g_B(x)$ with $\omega_B=0$ and possible shape modes with $0<\omega<m$.

Following Refs.~\cite{cahill76,mekink} it is convenient to decompose our real scalar field and its conjugate momentum not in terms of the usual operators $a^\dag$ and $a$ that create and destroy plane waves
\bea
\phi(x)&=&\pin{p}\frac{1}{\sqrt{2\omega_p}}\left(a^\dag_p+a_{-p}\right)e^{-ipx}
\nonumber\\
\pi(x)&=&i\pin{p}\sqrt{\frac{\omega_p}{2}}\left(a^\dag_p-a_{-p}\right)e^{-ipx} \label{osc}
\eea
but rather in terms of the fields $b^\dag_k$ and $b_k$ that create  and destroy normal modes, together with a zero-mode $\phi_0$ for the field and $\pi_0$ for its conjugate momentum.   Explicitly, $\phi(x)$ is decomposed
\bea
\phi(x)&=&\phi_0 g_B(x)+\int^+ \frac{dk}{2\pi} \frac{b^\dag_k g_k(x)+b_k g_k^*(x)}{\sqrt{2\ok{}}}\label{dec}\\
\pi(x)&=&\pi_0 g_B(x)+i\int^+ \frac{dk}{2\pi} \sqrt{\frac{\ok{}}{2}}\left(b^\dag_k g_k(x)-b_k g_k^*(x)\right).
 \nonumber
\eea
Here $\int^+ dk/2\pi$ is an integral over all real values of $k$, corresponding to continuum modes, and also a sum over imaginary values with nonzero frequencies, corresponding to shape modes. 

The plane wave and normal mode bases of the operator algebra are related by a Bogoliubov transformation.  It is possible to expand any operator in either basis.  The plane wave normal ordering $::_a$ is defined such that, when an operator is expanded in terms of $a^\dag$ and $a$, all $a$ appear on the right.  In the normal mode basis, one may define a normal ordering $::_b$ in which all $\pi_0$ and $b_k$ appear on the right.  There is a Wick's theorem \cite{mewick} which equates one normal ordering to the other plus a sum of contractions, each of which comes with the function $\I(x)$, defined to be the solution to
\beq
\partial_x \I(x)=\dpin{k}\frac{1}{2\omega_k}\partial_x\left|g_{k}(x)\right|^2 \label{di}
\eeq
that asymptotes to zero.  

One can evaluate $H\p$ \cite{mekink}
\bea
H\p&=&\df^\dag H\df=Q_0+\sum_{n=2}^{\infty}H_n\label{hesp}\\
H_{n}&=&\int dx \ch_{n}(x)\nonumber\\
\ch_{n(>2)}(x)&=&\frac{\V{n}}{n!} :\phi^n(x):_a\nonumber\\
H_2&=&Q_1+\frac{\pi_0^2}{2}+\pin{k}^+\omega_k b^\dag_k b_k\nonumber
\eea
where $Q_0$ and $Q_1$ are the tree-level and one-loop contributions to the kink mass
\bea
Q_0&=&\int dx f^{\prime 2}(x)\label{q0}\\
Q_1&=&-\frac{1}{4}\pin{p}\left[
\left|\tg_B(p)\right|^2\op\right.\nonumber\\
&&\left.
+\pin{k}^+\left|\tg_k(p)\right|^2
\frac{(\ok{}-\op)^2}{\op}
\right].
\nonumber
\eea
Here we have defined the inverse Fourier transforms
\beq
\tg(p)=\int dx g(x) e^{ipx}.
\eeq
Note that $H_n$ has $n/2-1$ powers of the coupling $\lambda$.  In Ref.~\cite{menormal}, diagrams were introduced which calculate masses of kink states.  Each line in these diagrams corresponds to a normal mode and, for each vertex which hosts $m$ loops and $n$ other lines ending at other vertices, $H_{2m+n}$ leads to a factor of
\beq
V_{\I\stackrel{m}{\cdots}\I,k_1\cdots k_n}\,=\,\int\, dx V^{(2m+n)}[\sqrt{\lambda}f(x)]\I^m(x)\prod_{i=1}^n  g_{k_i}(x)
.\label{vf}
\eeq

The two-loop correction to the energy of the kink ground state is \cite{me2looplett}
\bea
Q_2&=&\sum_{j=1}^5 Q_2^{(j)}\hsp
Q_2^{(1)}=\frac{V_{\I \I}}{8}\label{q2}\\
Q_2^{(2)}&=&-\frac{1}{8}\dpin{k}\frac{\left|V_{\I  k}\right|^2}{\ok{}^2}\nonumber\\
Q_2^{(3)}&=&-\frac{1}{48}\dpink{3} \frac{\left|V_{k_1k_2k_3}\right|^2}{\omega_{k_1}\omega_{k_2}\omega_{k_3}\left(\omega_{k_1}+\omega_{k_2}+\omega_{k_3}\right)}\nonumber\\
Q_2^{(4)}&=&\frac{1}{16}\dpink{2}\frac{\left|V_{Bk_1k_2}\right|^2}{\omega_{k_1}\omega_{k_2}\left(\ok1+\ok2\right)^2} \nonumber\\
Q_2^{(5)}&=& -\frac{1}{8}\dpin{k}\frac{\left|V_{BBk}\right|^2}{\ok{}^4}.
\nonumber
\eea

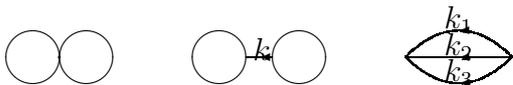
\begin{figure}
\setlength{\unitlength}{.35cm}
\begin{picture}(25,4)(-7,-1)
\put(-2,1){\circle{2}}
\put(-4,1){\circle{2}}
\put(-3.01,1){\circle*{.1}}

\put(6,1){\circle{2}}
\put(5,1){\vl{-1}0 {.5}}
\put(3,1){\circle{2}}
\put(4.6,1.3){\makebox(0,0){{$k$}}}

\qbezier(10,1)(12,3)(14,1)
\put(14,1){\vl{-1}0 2}
\put(12,2){\vector(-1,0) 0}
\put(12,0){\vector(-1,0) 0}
\qbezier(10,1)(12,-1)(14,1)
\put(12,2.4){\makebox(0,0){{$k_1$}}}
\put(12,1.4){\makebox(0,0){{$k_2$}}}
\put(12,0.4){\makebox(0,0){{$k_3$}}}

\end{picture}
\caption{The three diagrams respectively represent $Q_2^{(1)}$, $Q_2^{(2)}$ and $Q_2^{(3)}$. To obtain $Q_2^{(4)}$ $\left(Q_2^{(5)}\right)$, replace a normal mode with one (two) zero modes in the $Q_2^{(3)}$ diagram.}
\label{q2fig}
\end{figure}

The diagrams corresponding to these terms are shown in Fig.~\ref{q2fig}.  Eq.~(\ref{q2}) also yields the vacuum energy, if one fixes $f(x)$ to a minimum of $V$ and identifies $g_k$ with plane waves.  In that case only $Q_2^{(3)}$ is nonzero and in fact it is infinite, as $V_{k_1k_2k_3}\propto \delta(\kkk)$.  This divergence is to be expected, as at this order the vacuum has a finite energy density $\rv$.   The vacuum energy arises from the third diagram of Fig.~\ref{q2fig}, which is now equivalent to an ordinary Feynman diagram.  In the case of the kink, $V_{k_1k_2k_3}$ and so $Q_2^{(3)}$ has the same long distance divergence at $\kkk\rightarrow 0$ as the vacuum.  This may be expected as the continuum normal modes are plane waves far from the kink, and so, except for a phase shift, are not affected by the kink.   

Note the appearance of tadpole diagrams.  These do not vanish as a result of two-loop corrections to the expectation value of $\phi$.  We expect that we could remove these tadpoles by shifting $f(x)$ in $\df$ by a $\I(x)$-dependent loop correction, and that the shifted $f(x)$ may have a simple interpretation in terms of a Fourier tranformed form factor.  We leave this to future work, where we intend to investigate form factors more comprehensively.

$Q_2^{(3)}$ in (\ref{q2}) contains an integral over all continuum modes as well as a sum over discrete modes.  The discrete modes are bound states, which do not extend to spatial infinity and so do not contribute to the divergence.  Therefore if we decompose $Q_2^{(3)}$ into four summands $Q_2^{(3I)}$ which contain $I$ continuous modes and $3-I$ discrete modes, then the divergence will lie in
\beq
Q_2^{(33)}=-\frac{1}{48}\pink{3} \frac{\left|V_{k_1k_2k_3}\right|^2}{\omega_{k_1}\omega_{k_2}\omega_{k_3}\left(\omega_{k_1}+\omega_{k_2}+\omega_{k_3}\right)}.
\eeq

\section{Derivation of the Kink Mass}

We will now explain how this divergence can be removed.  Our strategy will be to set the two-loop vacuum energy to zero by adding a counterterm to the Hamiltonian density
\beq
\chct=-\rv.
\eeq
As $\rv$ is a $c$-number, it commutes with $\df$ and so also appears as is in the kink Hamiltonian $H\p$.  It is of order $O(\lambda)$ and so it contributes to $\ch_4$.  Now we will repeat the derivation of $Q_2^{(3)}$ in Ref.~\cite{me2loop}, modified to include this new counterterm.

The derivation begins with the eigenvalue equation (\ref{hp}) expanded using the semiclassical expansions (\ref{semi}) and (\ref{hesp})
\beq
\sum_{j=0}^i \left(H_{i+2-j}-Q_{\frac{i-j}{2}+1}\right)\vac_j=0.
\eeq
Here the index $i$ is the order of perturbation theory.  The equations at order $i=0$ and $i=1$
\beq
Q_1\vac_0=H_2\vac_0\hsp 0=H_3\vac_0+H_2\vac_1
\eeq
imply that the zeroeth order state $\vac_0$ satisfies
\beq
\pi_0\vac_0=b_k\vac_0=0
\eeq
and that the first order state $\vac_1$ has the following contribution from the three normal mode sector of the free Fock space
\beq
\vac_1\supset -\dpink{3}\frac{V_{k_1k_2k_3}}{6\sum_i^3 \ok{i}}\frac{b^\dag_{k_1}b^\dag_{k_2}b^\dag_{k_3}}{\sqrt{8\ok1\ok2\ok3}}\vac_0. \label{gameq}
\eeq

$Q_2$ is determined by the coefficient of $\vac_0$ of the order $i=2$ eigenvalue equation
\beq
Q_2\vac_0=H_4\vac_0+H_3\vac_1+(H_2-Q_1)\vac_2. \label{eom2}
\eeq
On the right hand side one can identify two contributions to $Q_2^{(3)}$: the contribution from $H_3$ acting on (\ref{gameq}), which leads to the term already in Eq.~(\ref{q2}), and also a new contribution arising from the counterterm acting on $\vac_0$
\bea
&&Q_2\vac_0\supset \int dx \left[-\rv\vac_0\right.\label{q2p}\\
&&\left.-\ch_3(x)  \dpink{3}\frac{V_{k_1k_2k_3}}{6\sum_i^3 \ok{i}}\frac{b^\dag_{k_1}b^\dag_{k_2}b^\dag_{k_3}}{\sqrt{8\ok1\ok2\ok3}}\vac_0\right]. \nonumber
\eea

Let
\beq
\sigma_{k_1k_2k_3}(x)=\V{3} g_{k_1}(x) g_{k_2}(x) g_{k_3}(x)
\eeq
which we note integrates to $V_{k_1k_2k_3}$.  Substituting  (\ref{dec}) into (\ref{hesp}), one sees that $\ch_3$ contains the term
\beq
\ch_{3}(x)\supset\frac{1}{6} \pink{3}\sigma_{k_1k_2k_3}(x)\frac{b_{-k_1}b_{-k_2}b_{-k_3}}{\sqrt{8\ok1\ok2\ok3}}.
\eeq
Note that the normal ordering is not relevant here, it only affects terms with contractions between $b$ and $b^\dag$.  Inserting this term into (\ref{q2p}) one finds the vacuum subtracted $Q_2^{(33)}$ 
\bea
\tilde{Q}_2^{(33)}&=&-\int dx \left[\rv \label{princ}\right.\\
 &+&\left.
\frac{1}{48}\pink{3} \frac{V_{k_1k_2k_3}\sigma_{-k_1-k_2-k_3}(x)}{\omega_{k_1}\omega_{k_2}\omega_{k_3}\left(\omega_{k_1}+\omega_{k_2}+\omega_{k_3}\right)}
\right].\nonumber
\eea
Our main result is that the kink mass $\tilde{Q}_2$ is obtained from the kink energy $Q_2$ in (\ref{q2}) by replacing $Q_2^{(33)}$, the integral over continuous normal modes in $Q_2^{(3)}$, with (\ref{princ}).

\section{Example: The $\phi^4$ Double Well}

Consider the $\phi^4$ double well, defined by 
\beq
V[\sqrt{\lambda}\phi(x)]=\frac{\lambda\phi^2}{4}\left(\sqrt{\lambda}\phi(x)-m\sqrt{2}\right)^2
\eeq
with classical kink solution
\beq
f(x)=\frac{m}{\sqrt{2\lambda}}\left(1+\tanh\left(\frac{m x}{2}\right)\right). \label{f}
\eeq
The normal modes are
\bea
g_k(x)&=&\frac{e^{-ikx}}{\ok{} \sqrt{m^2+4k^2}}\left[2k^2-m^2\right. \label{nmode}\\
&&\left.
+(3/2)m^2\sech^2\left(\frac{m x}{2}\right)-3im k\tanh\left(\frac{m x}{2}\right)\right]\nonumber\\
g_S(x)&=&-i\frac{\sqrt{3m}}{2}\tanh\left(\frac{m x}{2}\right)\sech\left(\frac{m x}{2}\right)\nonumber\\
g_B(x)&=&\frac{\sqrt{3m}}{2\sqrt{2}}\sech^2\left(\frac{m x}{2}\right)\nonumber
\eea
where $g_S$ corresponds to $k=im/2$ and so has frequency $\omega_S=m\sqrt{3}/2$.

The vacuum energy density at order $\lambda$ is \cite{serone}
\beq
\rv(x)=\left(\frac{1}{24}-\frac{\psi^{(1)} (1/3)}{16\pi^2}
\right)\lambda
\sim -0.0222644\lambda.
\eeq
The tree-level and one-loop contributions to the kink mass can be found from (\ref{q0})
\beq
Q_0=\frac{m}{3\lambda}
\hsp
Q_1=\left(-\frac{3}{2\pi}+\frac{1}{4\sqrt{3}}\right)m \label{q1}
\eeq
and agree, adjusting the convention for $m$ by the appropriate $\sqrt{2}$, with Ref.~\cite{dhn2}.

Using (\ref{q2}) and (\ref{princ}) we obtain \cite{phi4mass} the following contributions to the kink mass, including error bars only on dominant sources of error,
\bea
Q_2^{(1)}&\sim&  0.0152358 \frac{\lambda}{m}\hsp\,\,\,\,
Q_2^{(2)}\sim -0.0021398 \frac{\lambda}{m}\nonumber\\
Q_2^{(30)}&\sim& -0.0144574 \frac{\lambda}{m}\hsp\,\,\,
Q_2^{(31)}\sim -0.0017490 \frac{\lambda}{m}\nonumber\\
Q_2^{(32)}&\sim& -0.0623024(2) \frac{\lambda}{m}\hsp\,\,\,\,\,\,\,
\tilde{Q}_2^{(33)}\sim 0.145167(5) \frac{\lambda}{m}\nonumber\\
Q_2^{(4)}&\sim& 0.0076800 \frac{\lambda}{m}\hsp\,\,
Q_2^{(5)}=-0.075\frac{\lambda}{m}
\eea
which sum to the two-loop correction to the $\phi^4$ kink mass
\beq
\tilde{Q}_2\sim 0.012633(5)\frac{\lambda}{m}.
\eeq
This result is positive, which is consistent with the studies cited in the Introduction except for Ref.~\cite{latt09}.  However it is somewhat smaller than the best fit value of Refs.~\cite{slava,serone}, and in the later case lies just outside of the error bars.   Our result only contains the two-loop corrections, whereas the cited studies are either nonperturbative or higher order including Borel resummation, therefore it is meaningful to compare them only at weak coupling.  In this regime, our uncertainty is several orders of magnitude smaller than that of the cited studies.

\section* {Acknowledgement}

\noindent
JE is supported by the CAS Key Research Program of Frontier Sciences grant QYZDY-SSW-SLH006 and the NSFC MianShang grants 11875296 and 11675223.   JE also thanks the Recruitment Program of High-end Foreign Experts for support.


\begin{thebibliography}{99}

\bibitem{ssh}
W. Su, J. R. Schrieffer, A. J. Heeger, 
``Solitons in Polyacetylene,''
Phys. Rev. Lett. 42 25 (1979) 1698.

\bibitem{crystal}
Y. Wada, J. R. Schrieffer,  
``Brownian motion of a domain wall and the diffusion constants,''
 Phys. Rev. B \textbf{18}  (1978) no.8, 3897.

\bibitem{graphene}
R.~D.~Yamaletdinov, V.~A.~Slipko and Y.~V.~Pershin,
``Kinks and antikinks of buckled graphene: A testing ground for the $\phi^4$ field model,''
Phys. Rev. B \textbf{96} (2017) no.9, 094306
[arXiv:1705.10684 [cond-mat.mes-hall]].

\bibitem{dhn2}
  R.~F.~Dashen, B.~Hasslacher and A.~Neveu,
  ``Nonperturbative Methods and Extended Hadron Models in Field Theory 2. Two-Dimensional Models and Extended Hadrons,''
  Phys.\ Rev.\ D {\bf 10} (1974) 4130.

\bibitem{gj}
N.~Graham and R.~L.~Jaffe,
``Unambiguous one loop quantum energies of (1+1)-dimensional bosonic field configurations,''
Phys. Lett. B \textbf{435} (1998), 145-151
[arXiv:hep-th/9805150 [hep-th]].

\bibitem{vega}
H.~J.~de Vega,
``Two-Loop Quantum Corrections to the Soliton Mass in Two-Dimensional Scalar Field Theories,''
Nucl. Phys. B \textbf{115} (1976), 411-428

\bibitem{sl}
M.~Lowe and M.~Stone,
``A TWO LOOP CALCULATION ABOUT A QUANTUM MECHANICAL INSTANTON,''
Nucl. Phys. B \textbf{136} (1978), 177-188
doi:10.1016/0550-3213(78)90021-4

\bibitem{lat98}
A.~Ardekani and A.~G.~Williams,
``Lattice study of the kink soliton and the zero mode problem for phi**4 in two-dimensions,''
Austral. J. Phys. \textbf{52} (1999), 929-937
[arXiv:hep-lat/9811002 [hep-lat]].

\bibitem{latt09}
A.~Rajantie and D.~J.~Weir,
``Quantum kink and its excitations,''
JHEP \textbf{04} (2009), 068
[arXiv:0902.0367 [hep-lat]].

\bibitem{mussardo}
A.~Coser, M.~Beria, G.~P.~Brandino, R.~M.~Konik and G.~Mussardo,
``Truncated Conformal Space Approach for 2D Landau-Ginzburg Theories,''
J. Stat. Mech. \textbf{1412} (2014), P12010
[arXiv:1409.1494 [hep-th]].

\bibitem{slava}
S.~Rychkov and L.~G.~Vitale,
``Hamiltonian truncation study of the $\phi^4$ theory in two dimensions. II. The $\mathbb Z_2$ -broken phase and the Chang duality,''
Phys. Rev. D \textbf{93} (2016) no.6, 065014
[arXiv:1512.00493 [hep-th]].


\bibitem{bajnok}
Z.~Bajnok and M.~Lajer,
``Truncated Hilbert space approach to the 2d $\phi^{4}$ theory,''
JHEP \textbf{10} (2016), 050
[arXiv:1512.06901 [hep-th]].

\bibitem{vary03}
D.~Chakrabarti, A.~Harindranath, L.~Martinovic and J.~P.~Vary,
``Kinks in discrete light cone quantization,''
Phys. Lett. B \textbf{582} (2004), 196-202
[arXiv:hep-th/0309263 [hep-th]].


\bibitem{serone}
M.~Serone, G.~Spada and G.~Villadoro,
``$\lambda \phi_2^4$ theory \textemdash{} Part II. the broken phase beyond NNNN(NNNN)LO,''
JHEP \textbf{05} (2019), 047
[arXiv:1901.05023 [hep-th]].


\bibitem{me2looplett}
J.~Evslin and H.~Guo,
``Alternative to collective coordinates,''
Phys. Rev. D \textbf{103} (2021) no.4, L041701
[arXiv:2101.08028 [hep-th]].

\bibitem{force1}
H.~Weigel,
``Emerging Translational Variance: Vacuum Polarization Energy of the $\phi^6$ Kink,''
Adv. High Energy Phys. \textbf{2017} (2017), 1486912
[arXiv:1706.02657 [hep-th]].

\bibitem{force2}
I.~Takyi, M.~K.~Matfunjwa and H.~Weigel,
``Quantum corrections to solitons in the $\Phi^8$ model,''
Phys. Rev. D \textbf{102} (2020) no.11, 116004
doi:10.1103/PhysRevD.102.116004
[arXiv:2010.07182 [hep-th]].


\bibitem{cahill76}
K.~E.~Cahill, A.~Comtet and R.~J.~Glauber,
``Mass Formulas for Static Solitons,''
Phys. Lett. B \textbf{64} (1976), 283-285

\bibitem{mekink}
J.~Evslin,
``Manifestly Finite Derivation of the Quantum Kink Mass,''
JHEP \textbf{11} (2019), 161
[arXiv:1908.06710 [hep-th]].

\bibitem{mewick}
J.~Evslin,
``Normal Ordering Normal Modes,''
Eur. Phys. J. C \textbf{81} (2021) no.1, 92
[arXiv:2007.05741 [hep-th]].

\bibitem{menormal}
J.~Evslin and H.~Guo,
``Excited Kinks as Quantum States,''
[arXiv:2104.03612 [hep-th]].

\bibitem{me2loop}
J.~Evslin and H.~Guo,
``Two-Loop Scalar Kinks,''
Phys. Rev. D \textbf{103} (2021) no.12, 125011
doi:10.1103/PhysRevD.103.125011
[arXiv:2012.04912 [hep-th]].

\bibitem{phi4mass}
J.~Evslin,
``The $\phi^4$ Kink Mass at Two Loops,''
[arXiv:2104.07991 [hep-th]].

\end{thebibliography}
\end{document}